\documentclass[preprint,showpacs,showkeys,aps,prb]{revtex4}
\usepackage[dvips]{graphicx}

\begin{document}

\title{Time's Arrow for Shockwaves ; Bit-Reversible Lyapunov and
``Covariant'' Vectors ; Symmetry Breaking} 

\author{
Wm. G. Hoover and Carol G. Hoover               \\
Ruby Valley Research Institute                  \\
Highway Contract 60, Box 601                    \\
Ruby Valley, Nevada 89833                       \\
}

\date{\today}

\pacs{47.40.Nm, 02.60.-x}

\keywords{Shockwaves, time reversibility, bit reversibility, Lyapunov instability}

\vspace{0.1cm}

\begin{abstract}
Strong shockwaves generate entropy quickly and locally.  The Newton-Hamilton
equations of motion, which underly the dynamics, are perfectly time-reversible.
How do they generate the irreversible shock entropy?  What are the symptoms of
this irreversibility?  We investigate these questions using Levesque and Verlet's
{\it bit-reversible} algorithm.  In this way we can generate an entirely imaginary
past consistent with the irreversibility observed in the present.  We use Runge-Kutta
integration to analyze the local Lyapunov instability of nearby ``satellite''
trajectories.  From the forward and backward processes we
identify those particles most intimately connected with the irreversibility
described by the Second Law of Thermodynamics.  Despite the perfect time
symmetry of the particle trajectories, the fully-converged vectors associated
with the largest Lyapunov exponents, forward and backward in time, are qualitatively
different.  The vectors display a time-symmetry breaking equivalent to {\it Time's Arrow}.
That is, in autonomous Hamiltonian shockwaves the largest local Lyapunov exponents,
forward and backward in time, are quite different.

\end{abstract}

\maketitle

\section{Introduction and Goals}

Loschmidt's Paradox contrasts the microscopic {\it time-reversible} nature of Newtonian-Hamiltonian
mechanics with the macroscopic {\it irreversibility} described by the Second Law of Thermodynamics\cite{b1,b2,b3}. 
A way out of this Time-Reversibility Paradox has been charted through the use of Gaussian isokinetic
or Nos\'e-Hoover canonical thermostats to model nonequilibrium steady states such as shear flows and
heat flows. Both these thermostating approaches control the kinetic temperature with time-reversible
frictional forces of the form $\{ \ F_{thermal} \equiv -\zeta p \ \}$ .

With either Gaussian or Nos\'e-Hoover  mechanics, a {\it compressible} form of Liouville's Theorem holds :
$$
\{ \ \dot q = (p/m) \ ; \ \dot p = F(q) - \zeta p \ \} \ \longrightarrow
$$
$$
(d\ln f/dt) \equiv (\partial \ln f/\partial t) +\sum [ \ \dot q\cdot (\partial f/\partial q) + 
\dot p\cdot (\partial f/\partial p) \ ]/f =
$$
$$
-\sum [ \ (\partial \dot q/\partial q) +                                                       
         (\partial \dot p/\partial p) \ ] = \sum \zeta \ > 0 \ .
$$
In any stationary or time-periodic nonequilibrium state the sums here (over thermostated
degrees of freedom) must necessarily have a {\it nonnegative} average value of $(\dot f/f)$
to prevent instability.  This is because a persistent negative value (with decreasing probability
density) would correspond to an ever-increasing ultimately-divergent phase volume.  Thus the Second Law of
Thermodynamics becomes a {\it Theorem} when either Gaussian isokinetic or Nos\'e-Hoover mechanics
is used to generate such a nonequilibrium state.

What can be done to reconcile Newtonian-Hamiltonian mechanics with the Second Law in the
{\it absence} of thermostats?  {\it Shockwaves}, modeled with Levesque and Verlet's
``bit-reversible'' version of Newtonian mechanics\cite{b4,b5},
provide a possible way forward.  Despite the perfect time reversibility of the underlying
mechanics the shock propagation and compression dynamics should somehow be more ``likely'' 
than its time-reversed expanding image, in which entropy would decrease with the shock
running backwards.  The present work was inspired by Reference 5, and is an extension of our work
in Reference 6.  Here we investigate the reversibility problem through detailed analyses of the
largest local Lyapunov exponent and its associated vectors, one forward and one backward in time.  At every
phase-space point the forward and backward ``local'' exponents and the two associated vectors
depend on time.  A striking forward-backward symmetry breaking is the observed result.  This finding
suggests that purely Newtonian dynamics is enough for a clear demonstration of the Second Law of
Thermodynamics {\it even in the absence of thermostating forces}.

In the following Section we describe the basic shockwave geometry and its thermodynamic
interpretation.  Next, we detail a method for accurate determination of the Lyapunov
exponents with a ``bit-reversible'' reference trajectory and a Runge-Kutta satellite
trajectory.  Results from this hybrid  algorithm and 
the conclusions which we draw from them make up the final Sections of the paper.  An
Appendix responds to questions and comments raised by Pavel Kuptsov, Harald Posch, Franz
Waldner, and an anonymous referee.

\section{Shockwaves}

\begin{figure}
\vspace{1 cm}
\includegraphics[clip,height=6in]{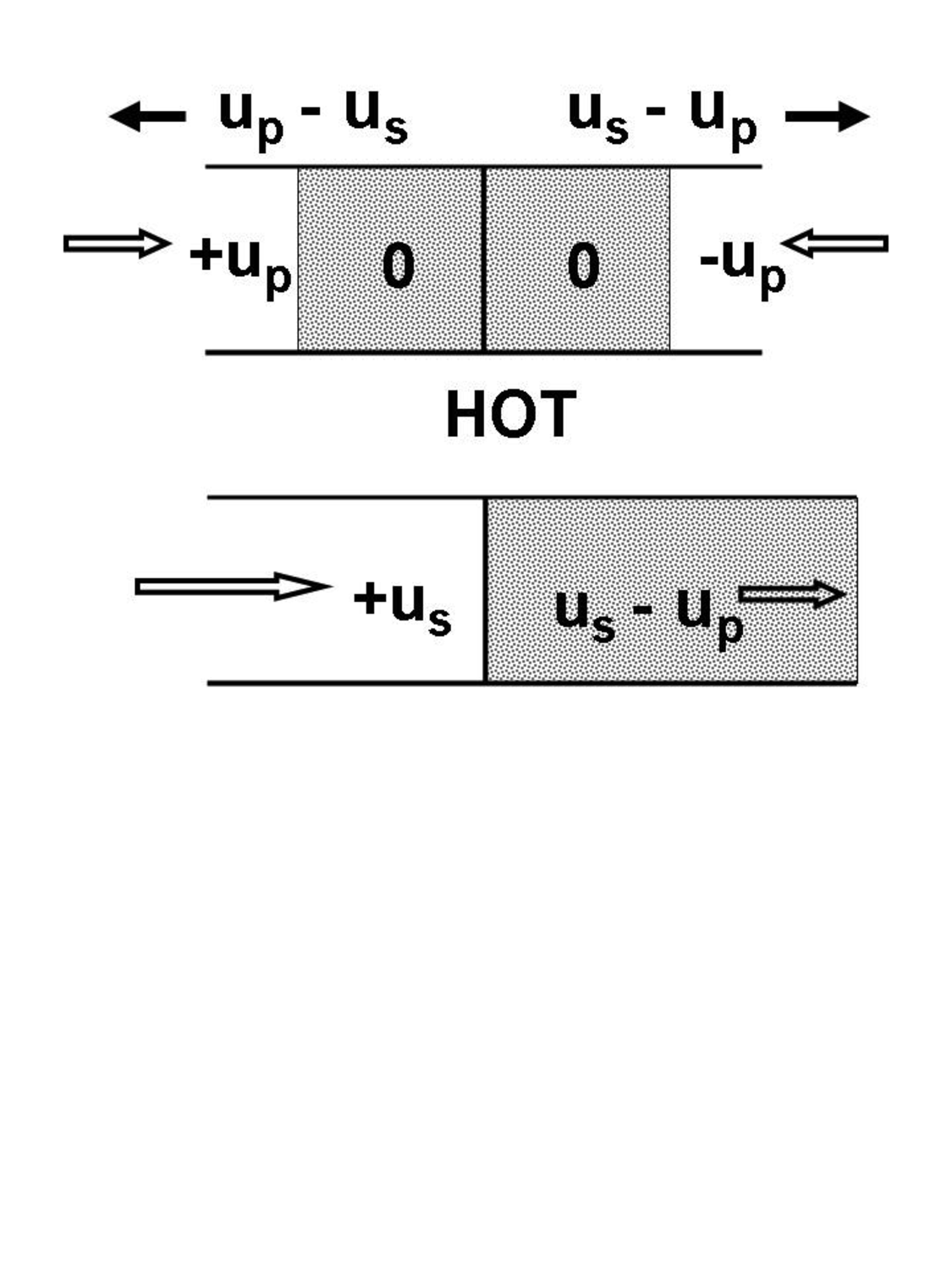}

\caption{
Shockwave generation.  In the upper view two mirror-image cold fluids collide with velocities
$\pm u_p$ generating a pair of shockwaves, moving at velocities $\pm(u_s - u_p)$ as they
convert the cold moving fluid to its hot (shaded grey) motionless shocked state.  The lower view shows
a stationary shockwave : cold fluid approaches the shockwave from the left, encounters the hot
slower fluid, and slows, $[ \ u_s \rightarrow u_s-u_p \ ]$ converting a portion of its kinetic
energy to internal energy.  The simulations discussed in the text and illustrated in the other
Figures correspond to the geometry of the upper view, with
periodic boundary conditions in both the $x$ and $y$ directions.
}
\end{figure}

Steady shockwaves are arguably the most irreversible thermodynamic processes possible.
In a typical shockwave cold low-pressure, low-temperature, low-entropy fluid is violently
{\it shocked} into a high-pressure, high-temperature, high-entropy state.  This conversion
takes place quickly, in just a few atomic collision times.  A high-speed collision of a right-moving cold fluid
with a fixed rigid wall generates a shockwave which decelerates the fluid from its initial
velocity, $u_p$ , to zero.  This inelastic collision with the wall generates a shockwave
moving to the left at speed $u_s - u_p$ .  The {\it force} on unit area of wall, $\rho u_su_p$ ,
is equal to the hot fluid {\it pressure}.  The initial kinetic energy of unit mass of fluid,
$(u_p^2/2)$ , becomes converted to the {\it internal energy} of the hot fluid.  At the same
time, the fluid {\it Entropy} necessarily {\it increases} because the shock process is patently
irreversible (from the thermodynamic standpoint).

In molecular dynamics simulations the collision of cold fluid with a wall is often modeled by
simulating the collision of a moving fluid with its mirror image.  An alternative approach models
a steady state with incoming cold fluid (from the left, at speed $u_s$) colliding with the slower hot fluid
(exiting to the right, at speed $u_s - u_p$).  See Figure 1 for these alternative views of the shock transition.  The
present work resembles the mirror approach.  In the next Section we consider a many-body system
with particles to the left of center moving right, at speed $u_p$ , and particles to the right
moving left, at the same speed.  As a result two shocks are generated at the interface.  They
move to the left and right with speed $u_s-u_p$ .  We emphasize that the dynamics here is ordinary
autonomous Hamiltonian (or Newtonian) mechanics, with no external forces and no time-dependent
boundary conditions.

Computer simulations of shockwaves with molecular dynamics, continuum mechanics, and the Boltzmann
equation have more than a 50-year history\cite{b2}.  Because the shock-compression process is nonlinear it
is natural to analyze it from the standpoint of chaos and nonlinear dynamics.  The present work is
devoted to exploring the microscopic mechanism for the shock process from the standpoint of
Lyapunov instability.  Some details of this work can be found in our recent ar$\chi$iv papers.
See also Chapter 6 of our book on Time Reversibility\cite{b2}.

\section{Integration Algorithm}

Levesque and Verlet\cite{b4} pointed out that an integration algorithm restricted to a
coarse-grained (or finite-precision) coordinate space can be made
precisely time-reversible provided that the coordinates $\{ \ q \ \}$ and
forces $\{ \ F(q) \ \}$ are rounded off
in a deterministic way.  Their algorithm is an {\it integer} version of the St\"ormer-Verlet
leapfrog algorithm :
$$
\{ \ q_+ - 2q_0 + q_- = (F_0\Delta t^2/m) \ \} \ .
$$
Both the lefthand and righthand sides are evaluated as (large, here nine-digit) integers.  It
is quite feasible to use 126-bit integers with the gfortran double-precision compiler.

Given the present and past values of the coordinates, $\{ \ q_0,q_- \ \}$, such an
integer algorithm can generate the future, or the past, for as many steps as one can afford. 
Accurate determinations of local instability require relatively long trajectories, as is
detailed in the Appendix.  It
is evident, for a finite phase space, that the entire history and future make up
(in principle) a single deterministic periodic orbit in the phase space.  Of course
the {\it time} required to achieve this periodicity can be tremendous.  The Birthday
Problem suggests that the Poincar\'e recurrence time is proportional to the square root of the
number of phase-space states, reaching the age of the Universe for a system size of
only a few atoms\cite{b2}

Our interest here is to examine the relative stability of the forward and backward
dynamics in order better to understand the
irreversible nature of the shockwave process.  Stability is best discussed in terms
of the Lyapunov instability in $\{ \ q,p \ \}$ phase space.  In order to use that
analysis for this integer-valued coordinate problem it is useful to {\it define} the
momentum $p$ at each timestep with an algorithm with a formal third-order accuracy :
$$
p_0 \equiv (4/3)\left[\frac{(q_+ - q_-)}{2\Delta t}\right] -
(1/3)\left[ \ \frac{(q_{++} - q_{--})}{4\Delta t} \ \right] \simeq
$$
$$
(4/3)[ \ \dot q + (\Delta t^2/6)(d^3q/dt^3) + (\Delta t^4/120)(d^5q/dt^5) \ ]
$$
$$
-(1/3)[ \ \dot q + (4\Delta t^2/6)(d^3q/dt^3) + (16\Delta t^4/120)(d^5q/dt^5) \ ]
$$
$$
= \dot q - (\Delta t^4/30)(d^5q/dt^5) \ .
$$
For simplicity, our particles have unit mass.

By keeping the Levesque-Verlet coordinates at five successive times ,
$$
\{ \ q_{--} \ , \ q_- \ , \ q_0 \ , \ q_+ \ , \ q_{++} \ \} \longrightarrow p_0 \ ,
$$
it is possible to
use the corresponding $\{ \ q,p \ \}$ states to  carry out accurate
Runge-Kutta integration forward (or backward) from $t_0$ to $t_{\pm \Delta t}$ so as
to find ``local'' (time-dependent) Lyapunov exponents.  To avoid programming errors it is best to start out
with a one-dimensional harmonic oscillator problem and to convert the resulting code to
deal with the shockwave problem.  In our implementation we have mapped both the coordinate
space and $[ \ (F\Delta t^2/m) \ ]$ onto the integer interval from $\{ \ 1 \ \dots 10^9 \ \} $.
The timestep we used was either 0.01 or 0.005 .  With these choices the truncation errors from
the nine-digit time integration are the same order as the equation of motion errors incurred by
the leapfrog algorithm.  This bit-reversible computer algorithm is exactly time-reversible\cite{b4}.

\section{Results}

We have carried out a variety of shockwave simulations, with different aspect ratios,
compression ratios, and forcelaws.  For definiteness we describe here calculations for the
twofold compression of a zero-pressure square lattice (nearest-neighbor spacing of unity
in two space dimensions) using the very smooth pair potential
$$
\phi(r<1) = (1 - r^2)^4 \ .
$$
Consider a $40\times 40$ periodic system with the leftmost half moving to the right
at speed 0.875 and the rightmost half moving to the left at the same speed.  We add random
thermal velocities corresponding to an initial temperature of order $10^{-5}$.  This arrangement
generates a pair of shockwaves traveling at approximately the same speed,
$u_s - u_p \simeq u_p = 0.875$ , to the right
and left.  These two shockwaves pass entirely through the sample in a time of about 20/1.75 ,
leaving behind a hot high-pressure fluid, compressed twofold and  occupying half the periodic container.

In order to study the effect of time reversal on the local stability of the shock propagation
process we reverse the time, ( $+\Delta t \rightarrow -\Delta t$ ) which reverses also the momenta and
the time ordering of the coordinates ,
$$
 \{ \ q_{--} \ , \ q_- \ , \ q_0 \ , \ q_+ \ , \ q_{++} \ \} \longrightarrow
 \{ \ q_{++} \ , \ q_+ \ , \ q_0 \ , \ q_- \ , \ q_{--} \ \} \ ,
$$
at equally-spaced intervals---see Figure 2---and verify that the resulting
time-periodic series of configurations is repeated to machine accuracy.  We verified also that
the results converge to a local limit independent of the time-reversal interval for sufficiently long
intervals.  We then compare the forward and backward Lyapunov exponents at corresponding times by
integrating a Runge-Kutta satellite trajectory constrained to lie within a fixed distance of the
bit-reversible reference trajectory.  We verified also that reducing the fixed distance corresponds
numerically to a well-defined limit.  Thus all the numerical work is straightforward and well-behaved.
\begin{figure}[h]
\vspace{1 cm}
\includegraphics[width=7.5cm,angle=-90]{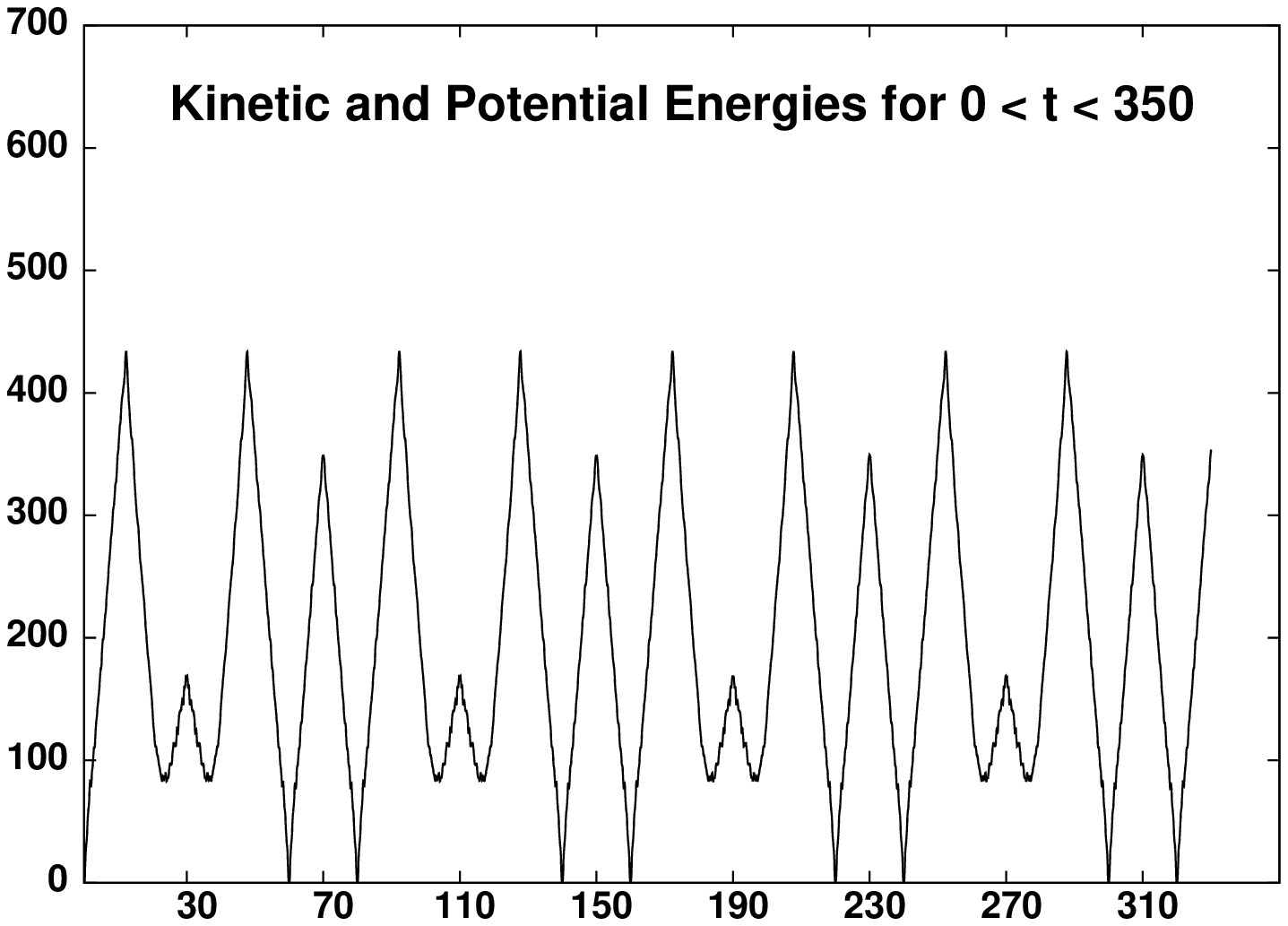}

\caption{
Kinetic (dashes) and Potential energies for a 1600-particle bit-reversible system as functions of time.
Maximum compression, about twofold, occurs at a time of 12.3 .  All the velocities are precisely
reversed at times of $\{ \ 30, 70, 110, 150, \dots \ \}$ so that maximum compression occurs
again at time $(60 - 12.3 = 47.7)$ .  The velocity reversal times correspond to the tick marks on
the abscissa.
}
\end{figure}

The forward and backward local Lyapunov exponents do converge to machine accuracy after a few
forward/backward integration cycles.  Each cycle begins and ends with a time reversal.  The
forward and backward exponents at corresponding times (same coordinates and opposite momenta)
turn out to be quite different.  Figure 2 shows the periodic behavior of
the kinetic and potential energies induced by the time reversals for this shockwave problem.
Figure 3 shows a portion of
the Lyapunov-exponent history for this same problem.  Note in the Figure that the
reversed trajectory briefly shows intervals with $\lambda _1$ negative (so that the satellite
trajectory shows an occasional tendency to approach the reference trajectory).

\begin{figure}[h]
\vspace{1 cm}
\includegraphics[height=10cm,width=7cm,angle=-90]{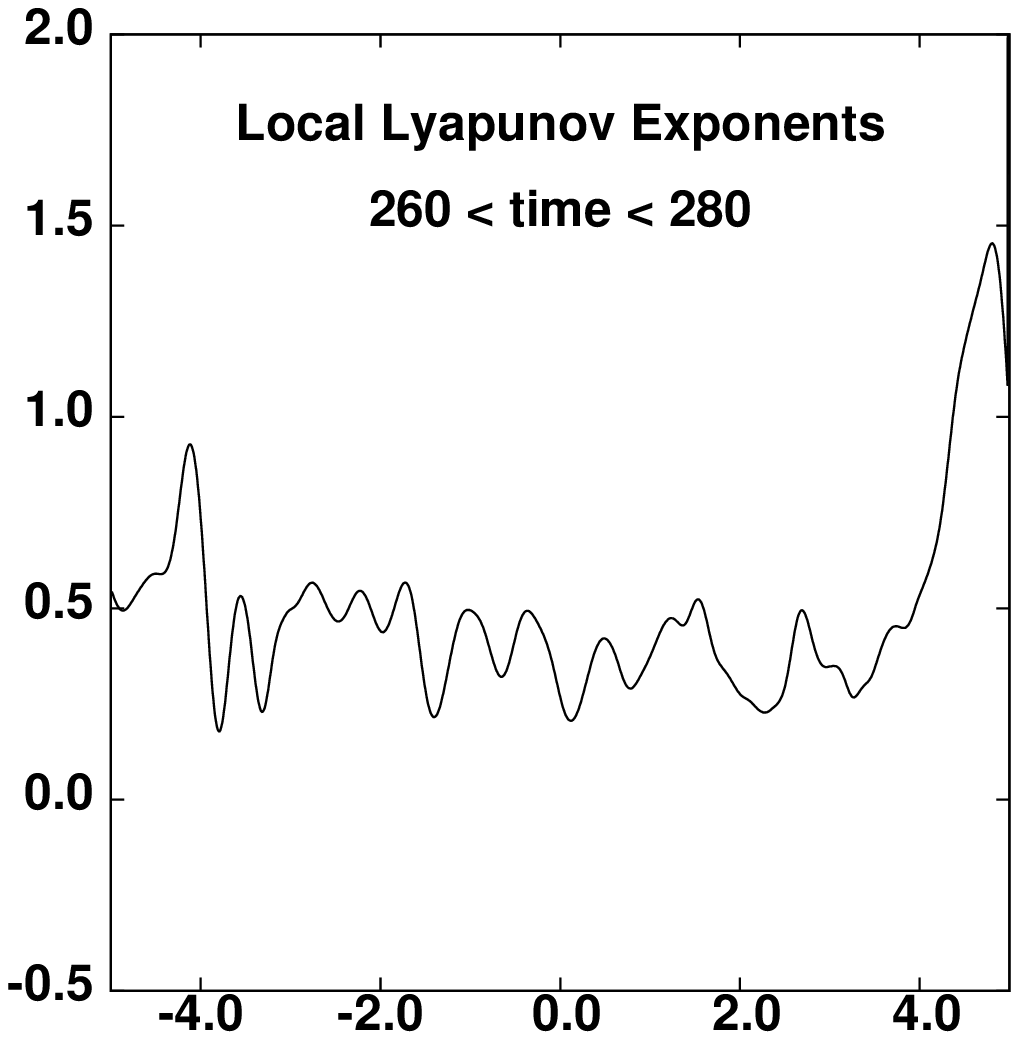}
\caption{
Largest Lyapunov exponents going forward in time (from 260 to 270) and going backward (dashes)
in time (from 280 to 270) are shown as two curves, at corresponding times. Note that the
 reversed exponent has brief episodes in which it is negative.
}
\end{figure}

The vectors corresponding to the forward and backward Lyapunov exponents are quite different.
For the problem described here the numbers of particles making above average contributions to
$\lambda _1^{\rm forward}$ and to $\lambda _1^{\rm backward}$ differ by roughly a factor of 2.
{\it Forward} in time the important particles are localised within the shocked material.
Backward in time there are more of these particles and they are distributed more nearly
homogeneously, as is shown in Figure 4 .  See also Reference 6 .

\newpage
\begin{figure}[h]
\vspace{1 cm}
\includegraphics[height=15cm,width=11cm,angle=-90]{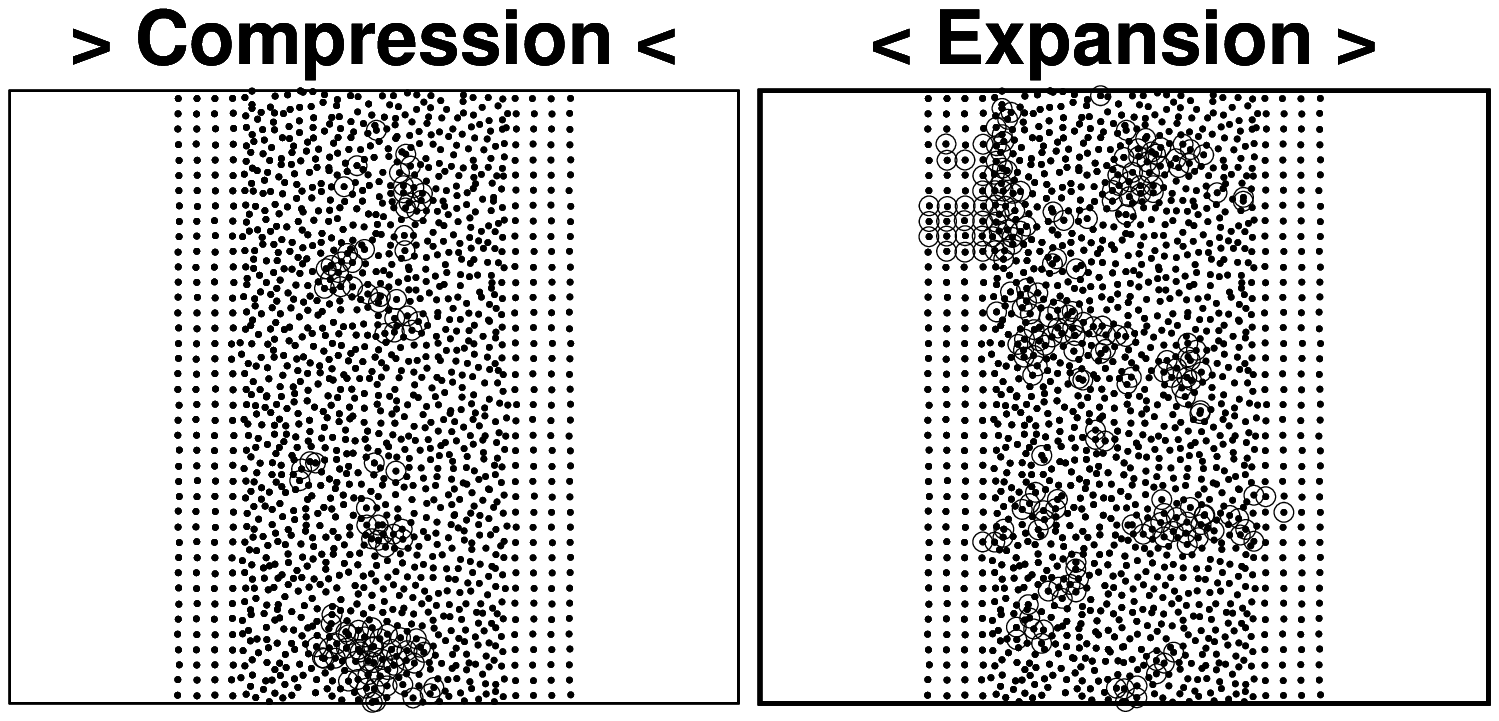}
\caption{ 
Particle positions at time 10.0 , after 1000 timesteps with $\Delta t = 0.01$ .  The motion is
precisely reversed at times of 30, 70, 110, ... . The kinetic energy reaches a minimum at a time of
12.3 , corresponding to maximum compression.  In the reversed expanding motion the minimum occurs
at a time of 47.7 .  After four periodic reference cycles, each of duration 80 , the nearby
satellite trajectory has converged to machine accuracy.  The larger open circles at the left
enclose the 101 particles making above-average contributions to $\lambda _1^{\rm forward}$ .
The more numerous particles (212) making above-average contributions in the
reversed motion are shown at the right with open circles.
}
\end{figure}

\section{Summary and Conclusions}

The results shown here indicate that despite the time-reversible and symplectic nature of Hamiltonian mechanics
there is a phase-space symmetry breaking (in the stability of sufficiently irreversible processes) which
distinguishes forward and backward trajectories.  In the forward-in-time shockwave problem the Lyapunov
instability is narrowly concentrated at the shockfronts.  In the reversed motion the instability
is much more widespread, throughout the expanding fluid.

These findings suggest that the dissipative Second Law of Thermodynamics has a purely-mechanical analog
and that Loschmidt's Paradox can be answered by symmetry breaking.  A quantitative relation linking the
Lyapunov vectors to entropy and irreversibility is still missing, but the present work suggests that
such a connection is a worthy goal.

Because the first ``covariant vector'' (in either time direction) and the corresponding exponents are
evidently identical to the first (most-positive) Gram-Schmidt exponents used here, these results hold also
in the ``covariant'' case.  The time-symmetry breaking found here is evidently generic for systems
sufficiently far from equilibrium.

\section{acknowledgment}

We thank Francesco Ginelli and Massimo Cencini for their invitation to contribute
this work to a special issue of the Journal of Physics A (Mathematical and Theoretical),
{\it Lyapunov Analysis from Dynamical Systems Theory to Applications} .  We appreciate
also a stimulating correspondence with Franz Waldner along with some helpful remarks
contributed by Pavel Kuptsov, Marc Mel\'endez, and Harald Posch.  Harald and an anonymous referee
both posed useful questions concerning this work which have substantially improved the text.  We address
these improvements further in the following Appendix.

\section{Appendix}

In mid-June of 2012 an anonymous referee requested a clear definition of the local Lyapunov exponents.
To provide it
consider a very long trajectory segment $r(t)$ with $-\tau < t < +\tau$.  No doubt a rigorous
approach would require that $\tau$ approach infinity.  Here $r(t)$ represents
the $4N$-dimensional phase-space trajectory of a (two-dimensional) $N$-body system. We denote it
$r_r(t)$, the ``reference'' trajectory.  Next consider a ``satellite'' trajectory constrained to
maintain a fixed small separation (infinitesimal in a rigorous approach) from the reference by a
Lagrange multiplier $\lambda$.  Hamilton's equations of motion, denoted here by $D$, govern both
the reference and the satellite trajectories :
$$
\dot r_r = D(r_r) \ ; \ \dot r_s = D(r_s) - \lambda(t)(r_s - r_r) \ .
$$
The additional Lagrange multiplier $\lambda(t)$, applied to the satellite trajectory, is chosen
to maintain the separation $|r_r - r_s|$ constant.

Suppose that the separation is small enough and that the time interval is long enough that $\lambda$ does
not depend upon the initial choice of $r_s(-\tau)$ . Then $\lambda(t=0)$ {\it is} the ``local''
Lyapunov exponent at time zero.  If the reference trajectory is stored and processed backward in
time, starting with a nearby satellite trajectory starting at $r_s(+\tau)$ the corresponding Lagrange
multiplier is the ``backward'' Lyapunov exponent.  It is worth pointing out that the trajectory
reversal can be carried out in either of two ways: (1) reverse the sign of $\Delta t$ in the integration
algorithm; or (2) reverse the momenta and the ordering of the coordinates.  The two approaches give
identical results.  

It is easy to show that reversing the sign of the Lagrange
multiplier along with the time gives $+\lambda^{\rm forward}=-\lambda^{\rm backward}$.  The exponent
measured invariably has a {\it positive} average value, indicating the tendency toward exponential
divergence of nearby trajectories.  For a short time it is usual to observe this simple sign change for
Hamiltonian systems.  We have found in the present work that if the past is sufficiently different to
the future then this symmetry is broken.  It is usual to check numerical results with finite precision
by using different algorithms with different timesteps.  We have done so in the present work.  Likewise
simulations with different offsets $|r_r - r_s|$ can be compared with results in which the equations
of motion for the satellite trajectory are evaluated by linearizing the reference motion equations
$D(r_r)$.  For an expanded discussion see References 2 and 7.
 
Harald Posch was bothered by the symmetry-breaking demonstrated here because the formal equations
for the local Lyapunov exponents are precisely time-reversible\cite{b7}.  This same objection could
be made for thermostated open systems.  In the case of steady-state open systems (thermostated
shear flows and heat flows are the simplest examples) the precisely time-reversible theory shows
that the phase-space volume associated with the flow must either shrink or expand on average.
Because expansion is ruled out (provided the occupied phase space is bounded) the flow must shrink,
with a negative sum of Lyapunov exponents.  Thus, for open systems, a bounded steady flow implies symmetry
breaking\cite{b8}.

For Hamiltonian systems (such as our shockwave problem) Liouville's Theorem implies a vanishing sum
of Lyapunov exponents :
$$
(d\ln f/dt) \equiv -\sum \lambda_i \equiv 0 \ .
$$
The sum includes all $4N$ exponents in the two-dimensional $N$-particle shockwave problem.  The symmetry
breaking demonstrated here closely resembles that found in open systems though there is not yet a formal
proof of this to provide necessary or sufficient conditions.

Accordingly, the reader may appreciate a physical analogy.  Imagine passengers in a speedy automobile, on a road
with long straightaways and occasional sharp curves.  The passengers are jostled by the curves and
recover on the straightaways.  In this analogy the reference trajectory corresponds to the auto and
the satellite trajectory to the passengers.  Thus the response of the satellite (to the past in
the usual forward time direction), can be quite different if the road is traveled in the opposite
(backward) direction.  The delayed response is reminiscent of Green and Kubo's linear-response theory.

Dr. Posch suggested to us that a satellite trajectory, just as the reference trajectory, could be
followed with a bit-reversible algorithm, ruling out any symmetry breaking.  The exponential growth
of perturbations rules out this approach. The time required for the convergence of the Lyapunov
vectors is considerably greater than $(1/\lambda_1)$.  For the shockwave problem the local Lyapunov
exponents converge visually in a time of order 10 and to machine accuracy in a time of a few hundred.

In our earlier work\cite{b6} on the shockwave model we simulated both the reference and the satellite
trajectories with Runge-Kutta integration.  The results were similar to those reported here. But
Runge-Kutta calculations could not be carried to much longer times (of order thousands) with confidence,
due to the irreversibility ( a local error of order $\Delta t^5$ ) of the Runge-Kutta algorithm.  The
present bit-reversible algorithm was developed in order to confirm that the results {\it do} converge as the
trajectory time between time reversals is increased.  The symmetry breaking is real and the results do
not depend upon the initial choice of the satellite trajectory.  The perfect time symmetry of the
bit-reversible reference trajectory makes it plain that the forward and backward local Lyapunov exponents
computed here  are equally valid descriptions of the trajectory's chaos.  Evidently these two choices
correspond to the two sets of adjoint covariant vectors recently described by Kuptsov and Parlitz\cite{b9}.

We felt the need for {\it simpler} Hamiltonian models to illustrate this novel symmetry breaking.
Accordingly we have carried out preliminary simulations, corroborating the symmetry breaking, by following
the motion of a two-dimensional anharmonic and chaotic diatomic molecule in a gravitational field and by
considering the collision of two relatively-small 37-particle drops.  We expect to report on the latter
simulations (for which we can characterize the complete local Lyapunov spectrum) in the very near future.  It is
tantalizing to imagine the insights into irreversible processes which the exploration of such Hamiltonian
symmetry-breaking will soon reveal.

We wish to point out a technical ``fly-in-the-ointment''.  Careful investigation reveals that the numerical
values of the local Lyapunov exponents have multifractal distributions\cite{b10,b11}.  Thus the exact
details of the local Lyapunov exponents do depend upon the chosen reference trajectory in a singular way.

As a postscript, the anonymous referee was still unsatisfied in January of 2013, leading us to publish this
work in Computational Methods in Science and Technology rather than the Journal of Physics A (which
had the manuscript under inconclusive review for 13 months, as of January 29, 2013, when we chose to withdraw
the manuscript).

\pagebreak

\end{document}